\documentclass[12pt]{article}
\usepackage{color,enumerate,amsmath,amssymb,xfrac,mathtools}
\def\hf{\sfrac{1}{2}}
\def\qu{\sfrac{1}{4}}

\def\he{\text{ht}}
\newcommand{\tbyt}[4]{\ensuremath{\left [ \begin{array}{cc} #1 & #2 \\ #3 & #4 \end{array} \right ]}}
\title{Re-imagining the Hardy-Weinberg Law} 
\author{YAP Von Bing ({\tt stayapvb@nus.edu.sg}) \\ Department of Statistics and Applied Probability \\ National University of Singapore}

\begin{document}

\maketitle



\section{Introduction} 

Suppose that a parental population has $k$ alleles $a_1,\ldots,a_k$ at an autosomal locus, and that the allele distributions in mothers and fathers are respectively $\{p_i^m\}$ and $\{p_i^f\}, i = 1,\ldots,k$. It is well-known that under random mating and the lack of mutation, selection or migration, allelic independence holds, i.e., the random maternal and paternal alleles, $M$ and $F$, are statistically independent, i.e., $\Pr(M=a_i,F=a_j) = p_i^m p_j^f$ for $1 \le i,j \le k$. Allelic independence is also known as random combination of gametes. Remarkably, the parental genotype distributions are irrelevant. Allelic independence leads easily to the Hardy-Weinberg Law. If the parental allele distributions are identical, $p_i^m = p_i^f = p_i, i = 1,\ldots,k$, then the offspring genotype distribution is
\begin{equation}
\label{hwe}
{\rm Pr}(a_ia_j) = \left \{ 
\begin{array}{ll} 
p_i^2, & i = j \ (\rm{homozygote}) \\
2p_ip_j, & i < j \ (\rm{heterozygote}) \\
\end{array} \right .
\end{equation} 
and this holds in subsequent generations produced under the same conditions. If the parental allele distributions differ, (\ref{hwe}) holds with $p_i = (p_i^m+p_i^f)/2$ in both female and male progenies. Thus, equilibrium is attained in at most two generations.

In 1908, Hardy and Weinberg independently proved (\ref{hwe}) in the case $k=2$, as is commonly presented in textbooks ([Ewe] pages 3--6, [Ham] pages 17-19). Edwards has a proof for arbitrary $k$ [Edw] (pages 6--7). All these arguments proceed by summing over and conditioning on relevant mating types. This article presents a simpler proof of allelic independence. A connection to Yule's paradox leads to a another proof using mating types, like the old approach, but neater. Furthermore, it is shown that allelic independence can hold under random mating and a certain form of fertility selection. Such combinations are completely characterised in the form of solutions to a homogeneous linear system of equations.
   
\section{The new proof}

Under random mating, a progeny comes about by three steps:
\begin{enumerate}
\item
Sampling of parents.
\item
Sampling of gametes, given parents.
\item
Fusion of gametes.
\end{enumerate}
Mendel's First Law combines steps 2 and 3 to obtain the genotype distribution from each mating type. Combining step 1 with the First Law necessitates summing over mating types, hence the algebraic complexity.  The simplicity in the new proof stems from first combining steps 1 and 2, which occur independently within the maternal and paternal populations.
\\

\noindent {\it Proof.} Suppose there are $n$ mothers. Put their alleles as rows in an $n \times 2$ matrix. Random mating means to choose a row at random, and by Mendel's First Law, an allele is chosen at random from this row. Hence, the maternal allele is chosen at random from the $2n$ alleles: $\Pr(M=a_i) = p_i^m, i=1,\ldots,k$. The process is analogous for the fathers, and the two processes are independent, so $\Pr(M=a_i,F=a_j) = p_i^mp_j^f, 1 \le i,j \le k$.
$\Box$

\section{Another proof along old lines}

The Yule's paradox [Yul] (or Simpson's paradox) refers to the phenomenon that relationships between variables in subgroups can be reversed when the subgroups are combined. In particular, two variables can be conditionally independent in all subgroups, but unconditionally dependent. From this perspective, allelic independence is a ``counter-example'': $M$ and $F$ are conditionally independent given any mating type, but they are independent. This is a clue to the existence of a simple proof. 

Let a square matrix $D$ represent a joint distribution, i.e., all entries are positive numbers summing to 1. Denote the $h$-th row sum as $d_{h+}$, and the $\ell$-th column sum as $d_{+\ell}$; these represent the marginal distributions. We say that $D$ is multiplicative if for every $h$ and $\ell$, $d_{h\ell} = d_{h+} d_{+\ell}$. Thus, $D$ is multiplicative if and only if the random variables are independent.

In the case of two alleles, consider the mating type $a_1a_1 \times a_1a_2$, i.e., the mother is $a_1a_1$ and the father is $a_1a_2$. By Mendel's First Law, $M$ must be $a_1$, while $F$ is equally likely to be $a_1$ or $a_2$. Their joint distribution is at row 1 and column 2 of Table 1, denoted by $J_{12}$. $J_{12}$ is multiplicative, so $M$ and $F$ are independent for this mating type. More generally, $M$ and $F$ are independent for every mating type, i.e., every matrix in Table 1 is multiplicative.
\begin{table}[h]
\begin{center}
\begin{tabular}{c|ccc}
 & $a_1a_1$ & $a_1a_2$ & $a_2a_2$ \\ \hline \\
 $a_1a_1$ & \tbyt{1}{0}{0}{0}  &  \tbyt{\hf}{\hf}{0}{0} & \tbyt{0}{1}{0}{0} \\ \\
 $a_1a_2$ & \tbyt{\hf}{0}{\hf}{0} & \tbyt{\qu}{\qu}{\qu}{\qu} & \tbyt{0}{\hf}{0}{\hf} \\ \\
 $a_2a_2$ & \tbyt{0}{0}{1}{0} & \tbyt{0}{0}{\hf}{\hf} & \tbyt{0}{0}{0}{1}
 \end{tabular}
\end{center}
\caption{Conditional distribution of $(M,F)$, given nine mating types. The matrix for $a_1a_1 \times a_1a_2$, at row 1 and column 2, is denoted $J_{12}$, etc.}
\end{table}

Suppose that the maternal and paternal genotype proportions are $\{u^m_{11},u^m_{12},u^m_{22}\}$ and $\{u^f_{11},u^f_{12},u^f_{22}\}$ respectively.  Under random mating, the probability that a female of genotype $h$ mates with a male of genotype $\ell$ is $w_{h\ell} = u^m_h u^f_\ell$, i.e., the weight matrix $W$ is multiplicative. Then the distribution of $(M,F)$ is given by the weighted average $J_W = \sum_{h,\ell} w_{h\ell} J_{h\ell}$. Clearly $J_W$ is multiplicative exactly if $M$ and $F$ are independent. Then allelic independence says $J_W$ is multiplicative whenever $W$ is multiplicative.

We make a key observation that leads to another proof of allelic independence.  The conditional distributions in Table 1 can be condensed, as shown in Table 2.  For example, $a_1a_1 \times a_1a_2$ gives two equally likely outcomes: $\{M=a_1,F=a_1\}$ and $\{M=a_1,F=a_2\}$, while $a_1a_2 \times a_1a_2$ gives four equally likely outcomes, etc.  For $k$ alleles, the condensed table is $g \times g$, where $g = k(k+1)/2$ is the number of genotypes.  
\begin{table}[h]
\begin{center} 
\begin{tabular}{c|ccc}
 & $a_1a_1$ & $a_1a_2$ & $a_2a_2$ \\ \hline
$a_1a_1$ & 1 & $\hf$ & 1 \\
$a_1a_2$ & $\hf$ & $\qu$ & $\hf$ \\
$a_2a_2$ & 1 & $\hf$ & 1
\end{tabular} \qquad
\begin{tabular}{c|cccccc}
 & $a_1a_1$ & $a_1a_2$ & $a_1a_3$ & $a_2a_2$ & $a_2a_3$ & $a_3a_3$ \\ \hline
$a_1a_1$ & 1 & $\hf$ & $\hf$ & 1 & $\hf$ & 1 \\
$a_1a_2$ & $\hf$ & $\qu$ & $\qu$ & $\hf$ & $\qu$ & $\hf$ \\
$a_1a_3$ & $\hf$ & $\qu$ & $\qu$ & $\hf$ & $\qu$ & $\hf$ \\
$a_2a_2$ & 1 & $\hf$ & $\hf$ & 1 & $\hf$ & 1 \\
$a_2a_3$ & $\hf$ & $\qu$ & $\qu$ & $\hf$ & $\qu$ & $\hf$ \\
$a_3a_3$ & 1 & $\hf$ & $\hf$ & 1 & $\hf$ & 1 \\
\end{tabular} 
\end{center}
\caption{Condensed conditional distributions of $(M,F)$ for two and three alleles.}
\end{table}

Remarkably, a further summary is available. Let ``hm'' and ``ht'' stand for ``homozygote genotype'' and ``heterozygote genotype'' respectively.  Then a $2 \times 2$ table suffices, indicating the probability of every relevant outcome from any mating type.   
\begin{table}[h]
\begin{center} 
\begin{tabular}{c|cc}
 & hm & ht \\ \hline
 hm & 1 & $\hf$ \\
 ht & $\hf$ & $\qu$
 \end{tabular} 
\end{center}
\caption{Summary of conditional distributions of $(M,F)$, for any number of alleles.}
\end{table}

Here is another proof of allelic independence using mating types. Let $\{u^m_h\}$ and $\{u^f_\ell\}$ be the parental genotype distributions, with corresponding allele distributions $\{p^m_i\}$ and $\{p^f_j\}$. Let ht$_1$ and ht$_2$ denote dummy heterozygote genotypes. Let $i$ and $j$ be fixed alleles, and let $\sum_{i \in \he_1}$ denote ``summing over all heterozygote genotypes containing $i$'', etc. Under random mating, Table 3 yields
\begin{eqnarray*}
\Pr(F=i,M=j) &=& u^m_{ii}u^f_{jj} \cdot 1 + 
\sum_{j \in \he_2} u^m_{ii} u^f_{\he_2} \cdot \frac{1}{2} + 
\sum_{i \in \he_1} u^m_{\he_1} u^f_{jj} \cdot \frac{1}{2} \ + \
\sum_{\mathclap{i \in \he_1, j \in \he_2}} u^m_{\he_1} u^f_{\he_2} \cdot \frac{1}{4} \\
&=& \left ( u^m_{ii} + \frac{1}{2} \sum_{i \in \he_1} u^m_{\he_1} \right )
\left ( u^f_{jj} + \frac{1}{2} \sum_{j \in \he_2} u^f_{\he_2} \right ) \\
&=& p^m_i p^f_j
\end{eqnarray*}

\section{Fertility selection}

Is there a set of non-multiplicative weights $W$ such that $J_W$ is multiplicative? The answer is yes. The weights in Table 4 are not multiplicative, but the distribution of $(M,F)$ is the same as that under random mating, if both parental genotype proportions are $\{\qu,\hf,\qu\}$.

\begin{table}[h]
\begin{center}
\begin{tabular}{c|ccc|c}
 & $a_1a_1$ & $a_1a_2$ & $a_2a_2$ & Row sum \\ \hline
 $a_1a_1$ & $\sfrac{3}{32}$ & $\sfrac{1}{16}$ & $\sfrac{3}{32}$ & $\qu$ \\
 $a_1a_2$ & $\sfrac{1}{16}$ & $\sfrac{3}{8}$ & $\sfrac{1}{16}$ & $\hf$ \\
 $a_2a_2$ & $\sfrac{3}{32}$ & $\sfrac{1}{16}$ & $\sfrac{3}{32}$ & $\qu$ \\ \hline
Column sum  & $\qu$ & $\hf$ & $\qu$ & 1
\end{tabular}
\end{center}
\caption{A set of non-multiplicative weights. For example, the proportion of $a_1a_1 \times a_1a_1$ is $\sfrac{3}{32} \ne \qu \times \qu$. But the associated joint distribution is multiplicative.}
\end{table}

In the case of $k$ alleles, there are $g = k(k+1)/2$ genotypes. Suppose $W$ is a $g \times g$ weight matrix, so that $J_W = \sum_{h,\ell} w_{h\ell} J_{h\ell}$ is a joint distribution of $(M,F)$. Let $W^*$ be the associated multiplicative weight matrix with $w^*_{h\ell} = w_{h+} w_{+\ell}$, hence $J_{W^*} = \sum_{h,\ell} w^*_{h\ell} J_{h\ell}$ is multiplicative. We seek to describe all $W$ such that $J_W = J_{W^*}$. In particular, for such $W$, $M$ and $F$ are independent. Define $S$ by
\[ W = W^* + S \]
Since the row and column sums of $W$ and $W^*$ are identical, those of $S$ are all 0:
\begin{equation}
\label{s_con0}
\sum_\ell s_{h\ell} = 0, \quad 1 \le h \le g, \qquad \sum_h s_{h\ell} = 0, \quad 1 \le \ell \le g 
\end{equation}
Now $J = J_{W^*}$ exactly when 
\begin{equation}
\label{s_con}
\sum_{h,\ell} s_{h\ell} J_{h\ell} = 0
\end{equation}
Conversely, given a multiplicative $W^*$, let $W = W^* + S$ be another weight matrix, with $S$ satisfying (\ref{s_con0}). Then $J_W = J_{W^*}$ if and only if (\ref{s_con}) holds. In the example,  
\[ W^* = \left [ \begin{array}{ccc} 
\sfrac{1}{16} & \sfrac{1}{8} & \sfrac{1}{16} \\ 
\sfrac{1}{8} & \sfrac{1}{4} & \sfrac{1}{8} \\ 
\sfrac{1}{16} & \sfrac{1}{8} & \sfrac{1}{16} \end{array} \right ], \qquad 
S = \left [ \begin{array}{ccc} 
\sfrac{1}{32} & \sfrac{-1}{16} & \sfrac{1}{32} \\ \sfrac{-1}{16} & \sfrac{1}{8} & \sfrac{-1}{16} \\ \sfrac{1}{32} & \sfrac{-1}{16} & \sfrac{1}{32} \end{array} \right ]\]
The entries of $S$ may be interpreted as fertility selection, i.e., frequency of progenies from mating type $h \times \ell$ increases by $s_{h\ell}$, from that under random mating, $w^*_{h\ell}$.  In the example, progenies of $a_1a_1 \times a_1a_2$, $a_1a_2 \times a_1a_1$, $a_1a_2 \times a_2a_2$ and $a_2a_2 \times a_1a_2$ decrease, while those of all other matings increase. 


We extract the following fact: \\

\noindent {\bf Theorem.} Let $\{u^m_h\}$ and $\{u^f_\ell\}$ be the genotype proportions of the maternal and paternal populations, and let $W^*$ be defined by $w^*_{h\ell} = u^m_h u^f_\ell$.  Let $S$ be a matrix satisfying (\ref{s_con0}) such that $W = W^*+S$ is positive. Assume random mating and fertility selection as described by $S$. If (\ref{s_con}) holds, then the joint distribution $\sum_{h,\ell} w_{h\ell} J_{h\ell}$ is the same as if there is no fertility selection; in particular, allelic independence holds. \\

The complete solution of (\ref{s_con0}) and (\ref{s_con}) is presented in the Appendix.  In the usual fertility selection [Pen], a fraction is multiplied to the mating type probability. Then rescaling the modified weights is necessary, resulting in some mating types becoming more and some less abundant than under random mating, much like our additive fertility selection.  Multiplicative modifier, a product of two factors depending on parental genotypes, was also studied [Bod].  While one could analogously represent $S$ as a sum, we note that under the constraint (\ref{s_con0}), the only such case is $S=0$.
In conclusion, allelic independence can arise from random mating without selection, or with fertility selection.  In particular, given random mating without mutation and migration, Hardy-Weinberg equilibrium does not imply no selection.

\vspace{.5cm}

\noindent {\it Acknowledgement.}  I thank Terry Speed and Anthony Edwards for valuable comments. \\

\noindent {\bf \large References} \\

\noindent [Bod] Bodmer, AF. Differential fertility in population genetics models, {\it Genetics} {\bf 51}: 411--424 (1965). \\
\noindent [Edw] Edwards, AWF. {\it Foundations of Mathematical Genetics 2e}, Cambridge University Press (2000). \\
\noindent [Ewe] Ewens, WJ. {\it Mathematical Population Genetics 2e}, Springer (2004). \\
\noindent [Ham] Hamilton, MB. {\it Population Genetics}, Wiley-Blackwell (2009). \\
\noindent [Har] Hardy, GH. Mendelian proportions in a mixed population, {\it Science} {\bf 28}:49--50 (1908). \\
\noindent [HC] Hartl, DL and Clark, AG. {\it Principles of Population Genetics 4e}, Sinauer Associates (2007). \\
\noindent [Pen] Penrose, LS. The meaning of ``fitness'' in human populations, {\it Annals of Eugenics} {\bf 14}: 301--304 (1949). \\
\noindent [Wei] Weinberg, W. \"{U}ber den Nachweis der Vererbung beim Menschen, {\it Jahreshefte des Verein f\"{u}r vaterl\"{a}ndische Naturkunde in W\"{u}rttemberg} {\bf 64}:368--382 (1908). \\
\noindent [Yul] Yule, GU. Notes on the theory of association of attributes in statistics. {\it Biometrika} {\bf 2}:121--134 (1903).

\section{Appendix}

We now present the complete characterisation of $S$ satisfying (\ref{s_con0}) and (\ref{s_con}), first in the biallelic case, then in general, followed by the symmetric case.      

\subsection{Two alleles: all solutions}

Writing
\[ S = \left [ \begin{array}{ccc}
s_{11,11} & s_{11,12} & s_{11,22} \\
s_{12,11} & s_{12,12} & s_{12,22} \\
s_{22,11} & s_{22,12} & s_{22,22} \end{array}
 \right ] \]
 (\ref{s_con}) is
\begin{eqnarray*}
s_{11,11} + \frac{1}{2}s_{11,12} + \frac{1}{2}s_{12,11} + \frac{1}{4}s_{12,12} = 0, &\qquad&
s_{11,22} + \frac{1}{2}s_{11,12} + \frac{1}{2}s_{12,22} + \frac{1}{4}s_{12,12} = 0, \\ 
s_{22,11} + \frac{1}{2}s_{22,12} + \frac{1}{2}s_{12,11} + \frac{1}{4}s_{12,12} = 0, &\qquad& 
s_{22,22} + \frac{1}{2}s_{22,12} + \frac{1}{2}s_{12,22} + \frac{1}{4}s_{12,12} = 0.
\end{eqnarray*}
It turns out that any one of the four equations suffices to determine $S$ completely. The solutions to the first top-left is a three-dimensional space.  Given a solution, the remaining five entries of $S$ are determined by (\ref{s_con0}). Now the other equations are automatically satisfied, which can be shown as follows. Using (\ref{s_con0}) on the first two rows of $S$, we have
\[ s_{11,11} + s_{11,12} + s_{11,22} = 0, \qquad \frac{1}{2}s_{12,11} + \frac{1}{2}s_{12,12} + \frac{1}{2}s_{12,22} =0 \]
whose sum equals the sum of the top two equations. Hence the top-right equation holds. Similarly, applying (\ref{s_con0}) to the first two columns of $S$ shows that the bottom-left equation holds. Finally, since the sum of the four equations is the sum of all entries of $S$, hence 0, the bottom-right equation holds. 

 \subsection{$k$ alleles: all solutions}

(\ref{s_con}) contains $k^2$ equations in $g^2$ unknowns, where $g = k(k+1)/2$ is the number of genotypes. The previous approach will be generalised. First, we establish that there are solutions to the $(k-1)^2$ equations with $1 \le i,j \le k-1$.  In general, simultaneous equations may be inconsistent, i.e., have no solutions; clearly this possibility does not arise when $k=2$. Note that the $(i,j)$-equation, $\sum_{h,\ell} s_{h\ell} J_{h\ell}(i,j) = 0$, has an unknown which does not appear in any other equation, namely $s_{ii,jj}$, because the mating type does not produce any other ordered genotype than $(i,j)$. Therefore the $(k-1)^2$ equations are consistent, and in fact, linearly independent. They involve only $s_{h\ell}$ where $h \ne kk$ or $\ell \ne kk$, i.e., the top left $(g-1) \times (g-1)$ submatrix of $S$. Hence the solutions are a subspace of dimension $(g-1)^2 - (k-1)^2$. Given such a solution, (\ref{s_con0}) determine the other unknowns. The second step is to check that consequently the equations for $(i,k), 1 \le i \le k-1$, $(k,j), 1 \le j \le k-1$ and $(k,k)$ hold. In the case $k=2$, this is accomplished by looking at certain rows or columns of $S$. The right generalisation is as follows. 

Let $1 \le i \le k-1$ be fixed. To show that the $(i,k)$-equation holds, i.e.,
\begin{equation}
\label{eq_ik}
s_{ii,kk} + \frac{1}{2} \sum_{j < k} s_{ii,jk} + \frac{1}{2} \sum_{i \in \he_1} s_{\he_1,kk} + \frac{1}{4} \sum_{i \in \he_1, j < k} s_{\he_1,jk} = 0
\end{equation}
we analyze the $(i,j)$-equation, which holds for $j < k$.
\begin{eqnarray*}
0 &=& s_{ii,jj} + 
\frac{1}{2} \sum_{j \in \he_2} s_{ii,\he_2} + 
\frac{1}{2} \sum_{i \in \he_1} s_{\he_1,jj} + 
\frac{1}{4} \sum_{i \in \he_1, j \in \he_2} s_{\he_1,\he_2} \\
&=& s_{ii,jj} + 
\left \{ \frac{1}{2} s_{ii,jk} + \frac{1}{2} \sum_{k \notin \he_2 \ni j} s_{ii,\he_2} \right \} + 
\frac{1}{2} \sum_{i \in \he_1} s_{\he_1,jj} +  
\left \{ \frac{1}{4} \sum_{i \in \he_1} s_{\he_1,jk} + \frac{1}{4} \sum_{i \in \he_1, k \notin \he_2 \ni j} s_{\he_1,\he_2} \right \} \\
\end{eqnarray*}
Each term in the first sum appears exactly twice among the $k-1$ equations. More specifically, if $1 \le j_1 < j_2 < k$, $s_{ii,j_1j_2}$ appears in the $(i,j_1)$- and $(i,j_2)$-equations but not in others. The same holds for the terms in the last sum. Hence summing the $k-1$ equations gives
\begin{equation}
\label{gen1}
\sum_{j < k} s_{ii,jj} + 
\frac{1}{2} \sum_{j < k} s_{ii,jk} + \sum_{j < k, k \notin \he_2 \ni j} s_{ii,\he_2} + 
\frac{1}{2} \sum_{i \in \he_1, j < k} s_{\he_1,jj} +
\frac{1}{4} \sum_{i \in \he_1, j < k} s_{\he_1,jk} + \frac{1}{2} \sum_{i \in \he_1, j < k, k \notin \he_2 \ni j} s_{\he_1,\he_2} = 0
\end{equation}
That the $ii$-row of $S$ sums to 0 can be written
\begin{equation}
\label{gen2}
s_{ii,kk} + \sum_{j < k} s_{ii,jj} + \sum_{j < k} s_{ii,jk} + \sum_{j < k, k \notin \he_2 \ni j} s_{ii,\he_2} = 0
\end{equation}
That any row of $S$ for a heterozygote genotype $\he_1$ containing $i$ sums to 0 can be written
\[ s_{\he_1,kk} + \sum_{j < k} s_{\he_1,jj} +  
\sum_{j < k} s_{\he_1,jk} + \sum_{j < k, k \notin \he_2 \ni j} s_{\he_1,\he_2} = 0 \] 
Summing over these heterozygote genotypes and multiplying by 1/2 gives
\begin{equation}
\label{gen3}
\frac{1}{2} \sum_{i \in \he_1} s_{\he_1,kk} + \frac{1}{2} \sum_{i \in \he_1, j < k} s_{\he_1,jj} +  
\frac{1}{2} \sum_{i \in \he_1, j < k} s_{\he_1,jk} + \frac{1}{2} \sum_{i \in \he_1, j < k, k \notin \he_2 \ni j} s_{\he_1,\he_2} = 0
\end{equation}
It can be readily checked that $(\ref{eq_ik}) + (\ref{gen1}) = (\ref{gen2}) + (\ref{gen3})$. Since (\ref{gen1},\ref{gen2},\ref{gen3}) hold, so does $(\ref{eq_ik})$, and this is valid for $1 \le i \le k-1$. Similarly, the $(k,j)$-equation holds for $1 \le j \le k-1$. An analogous argument as before shows that consequently the $(k,k)$-equation holds too. 

\subsection{$k$ alleles: symmetric solutions}

In general, fertility selection depends on the genotypes' parental origin, i.s., $s_{h\ell}$ need not equal $s_{\ell h}$. In cases where the origin does not matter, $S$ should be constrained to be symmetric. We now obtain all symmetric solutions. It is straightforward to verify that for any $h \ne \ell$, $J_{\ell h}$ is the transpose of  $J_{h\ell}$. It follows that
\[ \sum_{h,\ell} s_{h\ell} J_{h\ell}(i,j) = \sum_{h,\ell} s_{\ell h} J_{\ell h}(j,i) \]
so for $i \ne j$, the $(i,j)$- and $(j,i)$-equations are identical. Proceeding as in the previous subsection, a symmetric $(g-1) \times (g-1)$ submatrix of $S$ is first determined, which belongs to a subspace of dimension $(g-1)g/2 - (k-1)k/2$. Using (\ref{s_con0}), a symmetric $S$ is obtained. The same calculations show that $S$ indeed satisfies all $k^2$ equations. 

Now we present a particular symmetric solution which generalises the example in the Section 4. Suppose that the fertility selection coefficients are $\alpha$ for ho $\times$ ho, $\beta$ for ho $\times$ he or he $\times$ ho, and $\gamma$ for he $\times$ he, i.e., $S$ has the same block structure as Table 3.  All rows and columns of $S$ sum to 0 if and only if 
\[ \alpha + \frac{k-1}{2}\beta = 0, \qquad \beta + \frac{k-1}{2}\gamma = 0 \]
For $1 \le i, j \le k$, we have
\begin{eqnarray*}
 \sum_{h,\ell} s_{h\ell} J_{h\ell}(i,j) &=& \alpha \cdot 1 + \beta (k-1) \cdot \frac{1}{2} + \beta (k-1) \cdot \frac{1}{2} + \gamma (k-1)^2 \cdot \frac{1}{4} \\ 
 &=& \left \{ \alpha + \frac{k-1}{2}\beta \right \} + \frac{k-1}{2} \left \{ \beta + \frac{k-1}{2}\gamma \right \} \\
 &=& 0
 \end{eqnarray*}
 In the example, $\alpha = 1/32, \beta = -1/16, \gamma = 1/8$.

\end{document}